\documentclass[aps,prc,twocolumn,showpacs,nofootinbib]{revtex4-1}
\usepackage[utf8]{inputenc}
\usepackage{graphicx}   
\usepackage{latexsym}   
\usepackage{enumerate}
\usepackage{rotating,booktabs,multirow}
\usepackage{amsmath}
\usepackage{amsfonts}
\usepackage{amssymb}
\usepackage{multirow}
\usepackage{bm}
\usepackage{xcolor}
\usepackage{hyperref}

\begin{document}
\title{Investigating the cluster production mechanism with isospin triggering:\\ Thermal models versus coalescence models}

\author{Apiwit Kittiratpattana$^{1,3}$, Tom Reichert$^{1,2}$, Pengcheng~Li$^{4,5}$, Ayut Limphirat$^{3}$, Christoph Herold$^{3,*}$, Jan~Steinheimer$^{6}$, Marcus~Bleicher$^{1,2}$}

\affiliation{$^1$ Institut f\"ur Theoretische Physik, Goethe Universit\"at Frankfurt, Max-von-Laue-Strasse 1, D-60438 Frankfurt am Main, Germany}
\affiliation{$^2$ Helmholtz Research Academy Hesse for FAIR (HFHF), GSI Helmholtz Center for Heavy Ion Physics, Campus Frankfurt, Max-von-Laue-Str. 12, 60438 Frankfurt, Germany}
\affiliation{$^3$ Center of Excellence in High Energy Physics \& Astrophysics, School of Physics, Suranaree University of Technology, University Avenue 111, Nakhon Ratchasima 30000, Thailand}
\affiliation{$^4$ School of Nuclear Science and Technology, Lanzhou University, Lanzhou 730000, China}
\affiliation{$^5$ School of Science, Huzhou University, Huzhou 313000, China}
\affiliation{$^6$ Frankfurt Institute for Advanced Studies (FIAS), Ruth-Moufang-Str.1, D-60438 Frankfurt am Main, Germany}
\email{herold@g.sut.ac.th}

\begin{abstract}
Isospin triggering allows to distinguish coalescence from thermal production of light clusters in heavy ion collisions. Triggering on $Y(\pi^-) - Y(\pi^+)$ allows to select very neutron or proton rich final states. The deuteron (cluster) production with coalescence ($d \propto n \cdot p$) leads then to an inverse parabolic dependence of the deuteron yield on $\Delta Y_{\pi}$. In contrast, in a thermal model, cluster production is independent on $\Delta Y_{\pi}$. The observation of a maximum deuteron (cluster) yield as function of $\Delta Y_{\pi}$ provides confirmation of the coalescence mechanism.

\end{abstract}

\maketitle

\section{Introduction}
The exploration of the properties of matter governed by the theory of strong interaction (Quantum-Chromo-Dynamics, QCD) is a topic of highest interest. Ab-initio calculations based on lattice QCD methods have shown that such matter undergoes a transition at sufficiently high temperatures and/or baryonic densities \cite{Borsanyi:2010bp,Bazavov:2011nk}. Such temperatures are e.g. reached in accelerator facilities like the CERN-LHC, BNL-RHIC or CERN-SPS. At the high density frontier, laboratory experiments are performed at RHIC in the beam energy scan program, at GSI's SIS18 accelerator or at the future FAIR facility. In nature the high temperature transition from the deconfined Quark-Gluon-Plasma state to a hadronic system happened approximately a few microseconds after the Big Bang, while the high density regime is probed by neutron stars and neutron star mergers. Especially neutron star mergers have renewed the interest in the equation-of-state of nuclear matter at highest densities \cite{Most:2022wgo} because gravitational wave measurements might allow to pin down the equation-of-state of QCD matter with very high precision \cite{LIGOScientific:2018cki}. 

A central tool that is often used to infer the properties of the created matter are light clusters, e.g. deuterons, tritons and helium. For the production of such states one uses generally two complementary approaches: The statistical (thermal) model \cite{Cleymans:1992zc,Becattini:1997ii,Florkowski:2001fp,Cleymans:2005xv,Andronic:2010qu,Petran:2013lja,Vovchenko:2015idt,Andronic:2018qqt,Vovchenko:2019pjl} or coalescence \cite{Aichelin:1991xy,Nagle:1994wj,Bleicher:1995dw,Mattiello:1996gq,Nagle:1996vp,Puri:1996qv,Puri:1998te,Chen:2003ava,Ko:2010zza,Glassel:2021rod,Zhu:2015voa,Sombun:2018yqh,Sun:2020uoj,Gaebel:2020wid,Zhao:2021dka,Kireyeu:2022qmv,Kittiratpattana:2020daw,Kittiratpattana:2022knq}.
While both models provide similar results \cite{Mrowczynski:2016xqm,Sombun:2018yqh} over a wide range of collision energies they are very different in their physics assumptions: 
\begin{itemize}
    \item[(I)] The thermal model assumes the creation of a fully thermalized (mostly assumed grand canonical) fireball, which means that the clusters are produced at the chemical freeze-out at a temperature of 60-150 MeV (depending on the collision energies probed in large systems like Au+Au or Pb+Pb) from $\sqrt{s_\mathrm{NN}}=2.4-13000$ GeV. An often discussed problem with this model is the fact that lightly bound clusters may not form or survive in such a hot environment. This is also known from Big Bang nucleosynthesis under the term of deuteron bottleneck, which means that deuterons and higher mass light elements can only be formed if the temperature is on the order or below the binding energy of a few MeV. 
    
    Even within the thermal model itself such a tension is visible in certain energy ranges \cite{Harabasz:2022rdt}, and the clusters are often removed from the thermal fitting as they worsen the quality of the thermal fit significantly \cite{Motornenko:2021nds}.

    \item[(II)] In contrast, the coalescence model assumes that light clusters are formed at kinetic freeze-out, i.e. after the last collisions/decays have ceased and the system reaches the free-streaming regime. Here the formation is possible due to lower temperatures and due to the fact that no further collisions will destroy the formed cluster. It is clear that at the earlier chemical freeze-out, the temperature is higher and the volume of the source is smaller than at the later kinetic freeze-out where the temperature is lower and the volume is larger \cite{Reichert:2020yhx}. 

\end{itemize}

\begin{figure}
    \centering
    \includegraphics[width=\columnwidth]{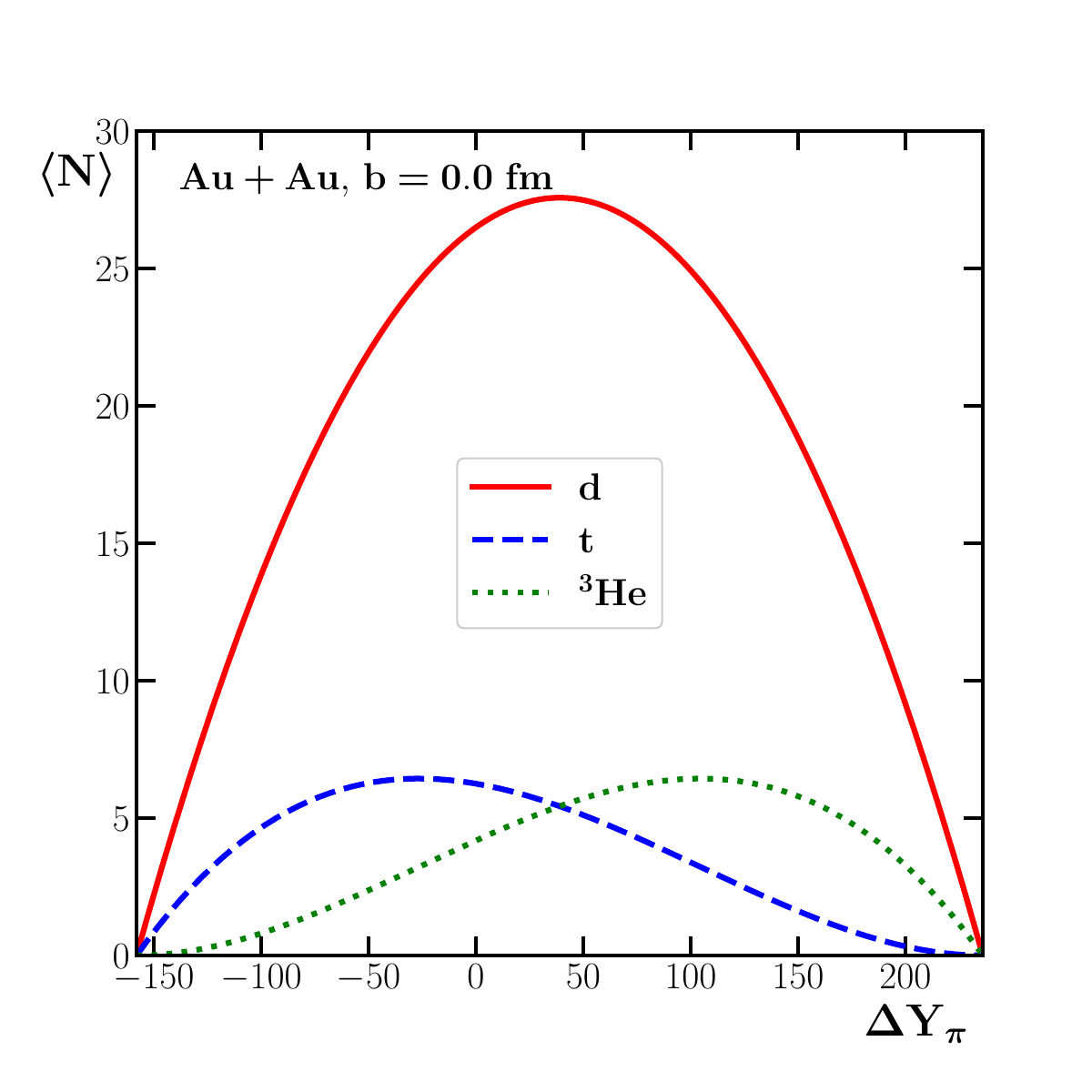}
    \caption{[Color online] Estimates for the deuteron (red full line), triton (blue dashed line) and ${}^3$He (green dotted line) production in Au+Au reactions as a function of $\Delta Y_{\pi}$.}
    \label{fig:eq_d_vs_deltapi}
\end{figure}

\begin{figure}
    \centering
    \includegraphics[width=\columnwidth]{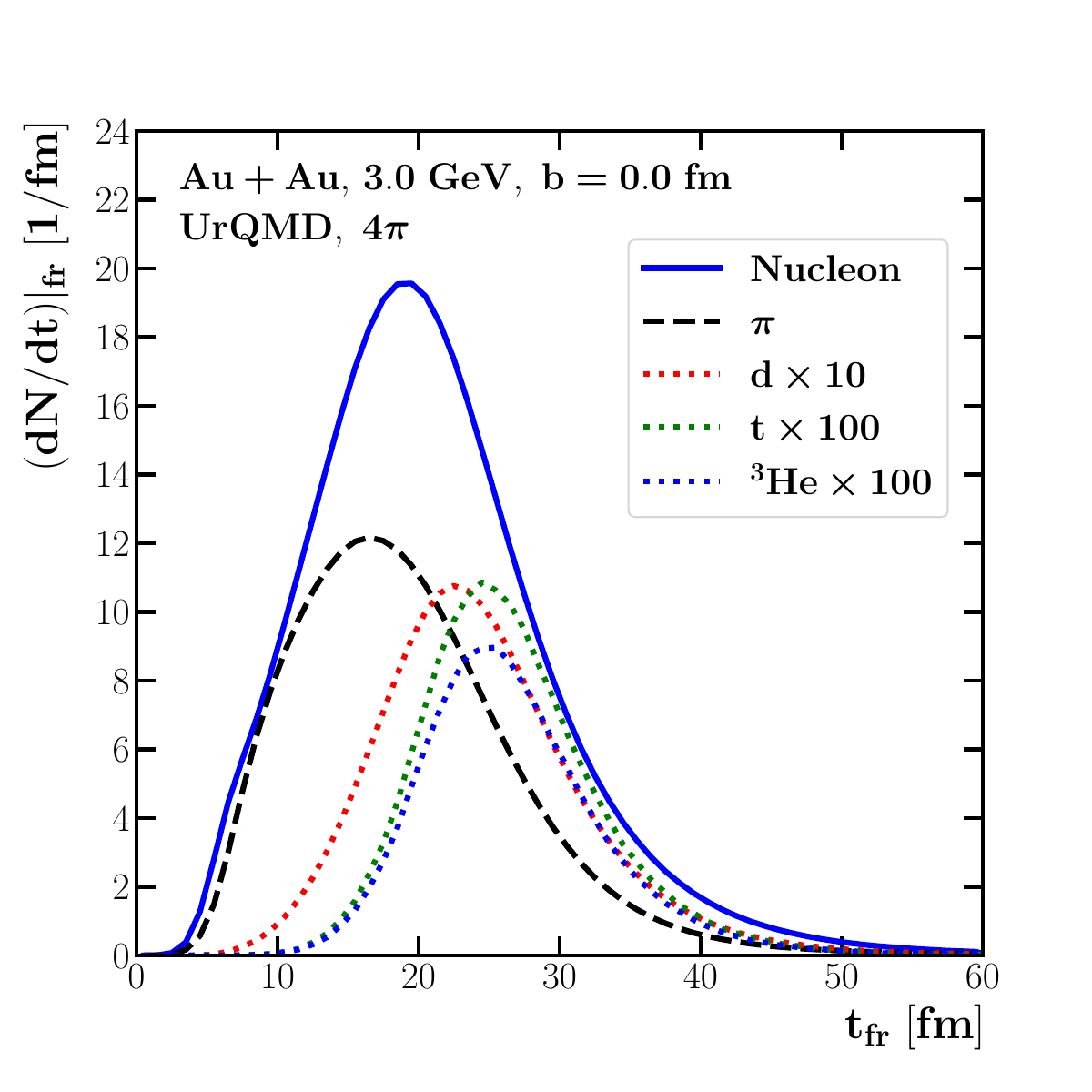}
    \caption{[Color online] Freeze-out time distribution of nucleons (full blue line), pions (dashed black line), deuterons (dotted red line), tritons (dotted green line) and ${}^3$He (dotted blue line).}
    \label{fig:freeze-out_time}
\end{figure}

Up to now it has not been possible to distinguish between both methods for cluster production, because the results for the mean values have been shown to be similar in the thermal model \cite{Vovchenko:2020dmv} and the coalescence approaches \cite{Hillmann:2021zgj}. Here, we propose a new method based on the isospin fluctuations that allows to distinguish thermal cluster production from the coalescence model.

The idea is the following: In the coalescence model, the deuteron production is proportional to the product of the number (densities) of the protons and neutrons at kinetic freeze-out. The number of protons and neutrons at kinetic freeze-out is related to the amount of emitted charged pions\footnote{It is clear that also other charged particles, e.g. Kaons can be produced. However, their yield is small and for the present argument it is sufficient to restrict the discussion to pions.} due to isospin conservation. E.g., assuming a fixed volume
\footnote{To obtain a first estimate of the effect, we assume that the number of participants and the $N/Z$ ratio in the participant nucleons does not fluctuate. In that sense, we speak of a fixed volume. In line with Ref. \cite{Kittiratpattana:2022knq} we use these primordial $N$ and $Z$ values for the estimates of the deuteron (and higher mass cluster) yields. Of course in a realistic situation this is not exactly true which is why we contrast our simple model estimates in the following with a detailed microscopic simulation of the UrQMD model in which no such assumptions are made and which shows almost the same behavior as our simplified model.}
and no production of (charged) pions, the ratio of neutrons to protons at kinetic freeze-out is equal to the initial ratio of neutrons to protons $N_\mathrm{fr}/Z_\mathrm{fr}=\alpha=N_\mathrm{Au}/Z_\mathrm{Au}$. The deuteron yield $d$ in the coalescence model is then proportional to the product of the neutron and proton numbers (or densities) $d\propto N_\mathrm{fr}\cdot Z_\mathrm{fr} = \alpha Z_\mathrm{fr}\cdot Z_\mathrm{fr}$. 

Let us now consider the case of charged pion production. At low energies, pions are produced during the collision, however travel inside the $\Delta$ resonance until the $\Delta$ escapes from the fireball and decays into a nucleon and a pion (kinetic freeze-out) \cite{Reichert:2019lny,Reichert:2020uxs}. The production of pions leads on average to an equipartion of isospin in the nucleon system. However, the decay of the $\Delta$'s is stochastic (given by the branching ratios) and introduces isospin fluctuations in the nucleon system at freeze-out. Let us look at the most extreme case: Initially the system in Au+Au reactions consisted of $2\cdot 79=158$ protons, triggering on an event with 158 emitted $\pi^+$ and no $\pi^-$ would lead to a nucleonic system at freeze-out consisting only of neutrons. In this case the production probability of a deuteron vanishes in the coalescence approach because there is no proton left to form a deuteron. The maximal deuteron yield is obviously reached when the pion emission created a system with $N_\mathrm{fr}=Z_\mathrm{fr}$. In summary, the emission of positive and negative pions changes the $N/Z$ ratio at freeze-out. This modified $N/Z$ ratio enters as an input into the coalescence model. Thus, the deuteron yield in the coalescence model depends on $\Delta Y_{\pi} = Y(\pi^-) - Y(\pi^+)$ and shows a distinct maximum. Analog arguments lead to maxima for higher mass clusters.

In case of the thermal model the situation is completely different \cite{Cleymans:1992zc,Becattini:1997ii,Florkowski:2001fp,Cleymans:2005xv,Andronic:2010qu,Petran:2013lja,Vovchenko:2015idt,Andronic:2018qqt,Vovchenko:2019pjl}. In a grand canonical model only the average positive and negative pion numbers are fixed, at chemical freeze-out, by an isospin chemical potential obtained from the initial $N/Z$ ratio (again we assume a fixed volume). 
The independent fluctuation of the positively and negatively charged pions lead to fluctuations of $\Delta Y_{\pi}$ also in the thermal model. However, these grand canonical $\Delta Y_{\pi}$ fluctuations do not influence the yield of the deuterons, because all hadron species fluctuate independent of each other in this approach. Even employing a fully canonical thermal model \cite{Vovchenko:2018fiy}, where the total isospin is fixed at chemical freeze out, does not change the situation much since the isospin is fixed before the decays of the resonances into pions. Since resonance decays are stochastic, even in the canonical thermal model the isospin content in the nucleons before decays (related to deuteron production) is not correlated with the isospin fluctuations $\Delta Y_{\pi}$ in the pions after decay.
Therefore, in a statistical thermal model, the deuteron yield should not depend on $\Delta Y_{\pi} = Y(\pi^-) - Y(\pi^+)$, it is flat as function of $\Delta Y_{\pi}$. The same argument holds for higher mass clusters.

We propose that the appearance of a local maximum of the deuteron yield at fixed (or at least tightly constrained) $A_\mathrm{part}$ as a function of $\Delta Y_{\pi}$  allows to distinguish thermal deuteron production from the coalescence approach. In addition higher mass clusters can be used to confirm this scenario.

\begin{figure}
    \centering
    \includegraphics[width=\columnwidth]{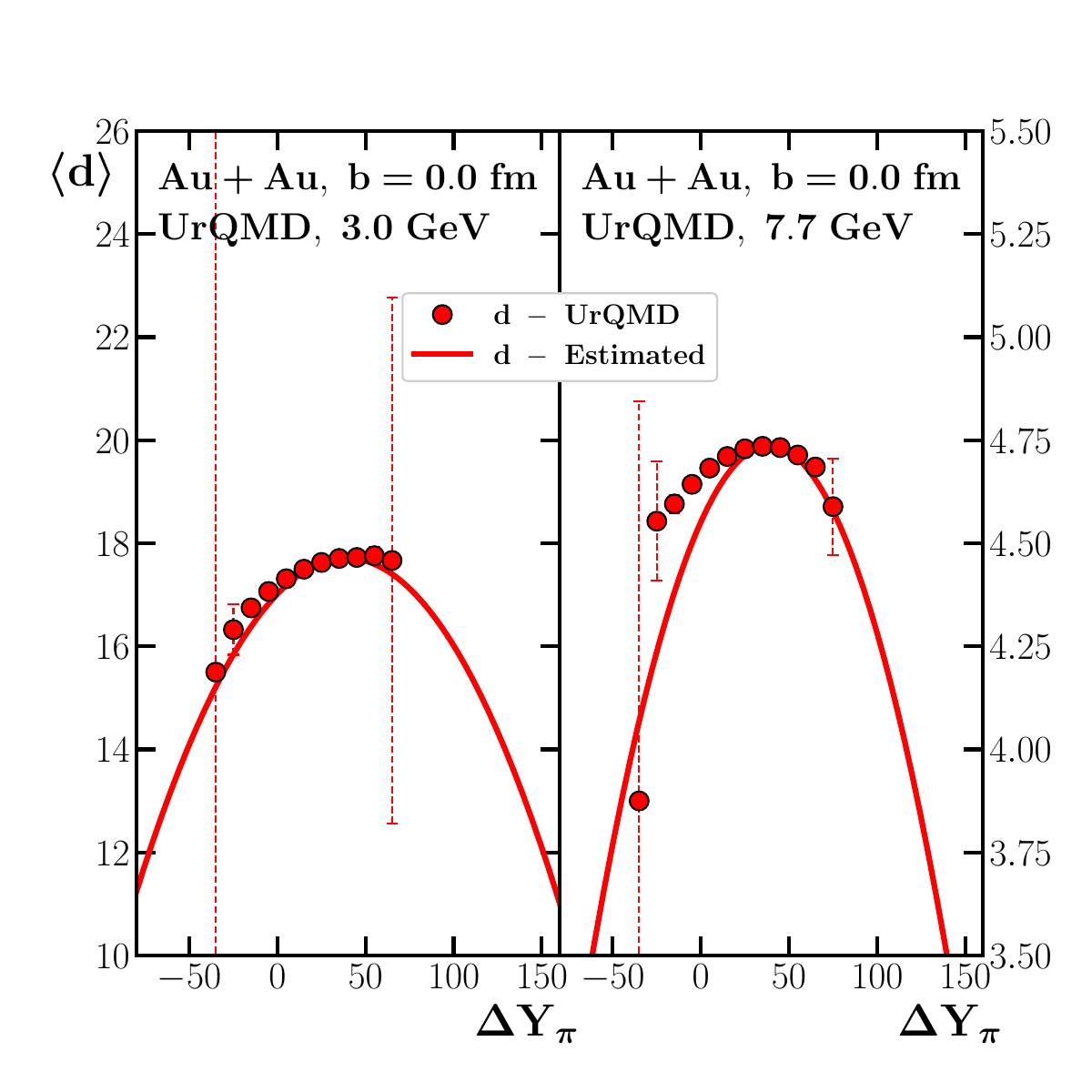}
    \caption{[Color online] Deuteron yield as a function of $\Delta Y_{\pi}$ for Au+Au reactions. The UrQMD results are shown by red circles. The estimated yield, Eq. \ref{eq:d}, is represented by the full red line. Left: Results at $\sqrt{s_\mathrm{NN}}=3$ GeV. Right: Results at  $\sqrt{s_\mathrm{NN}}=7.7$ GeV. }
    \label{fig:urqmd_d_vs_deltapi}
\end{figure}

\begin{figure}
    \centering
    \includegraphics[width=\columnwidth]{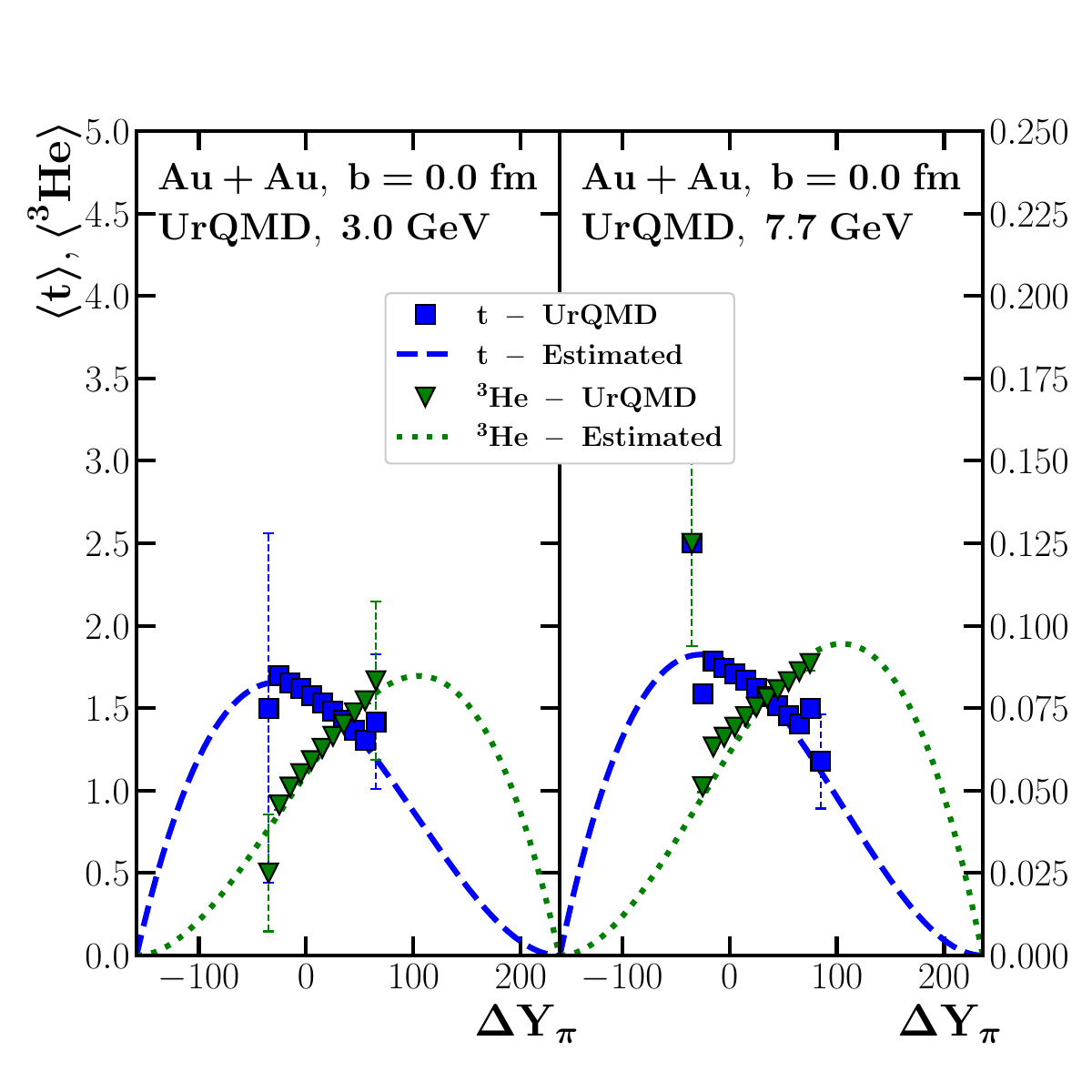}
    \caption{[Color online] Triton (blue squares and dashed blue line) and ${}^3$He (green triangles and dotted green line) yields  as a function of $\Delta Y_{\pi}$ for Au+Au reactions. The UrQMD results are shown by symbols. The estimated yields, Eqs. \ref{eq:t} and \ref{eq:He}, are represented by the lines. Left: Results at $\sqrt{s_\mathrm{NN}}=3$ GeV. Right: Results at  $\sqrt{s_\mathrm{NN}}=7.7$ GeV.}
    \label{fig:urqmd_t_he_vs_deltapi}
\end{figure}

\section{Simple estimates}
To illustrate the effect and its size, let us start from a simple model for the cluster production in Au+Au reactions. The yield of light clusters is related to the neutron and proton numbers at kinetic freeze-out. These are in turn related to the initial proton, $Z_\mathrm{Au}$, and neutron, $N_\mathrm{Au}$, numbers of the gold nucleus and to the $\Delta Y_{\pi} = Y(\pi^-) - Y(\pi^+)$ via (here $\Tilde{B}_A$ is some coalescence factor) \cite{Kittiratpattana:2022knq}:
\begin{equation}
    d = \Tilde{B}_2 \cdot (2N_{\rm Au} - \Delta Y_{\pi}) \cdot (2Z_{\rm Au} + \Delta Y_{\pi}) \label{eq:d}
\end{equation}

\begin{equation}
    t = \Tilde{B}_3 \cdot (2N_{\rm Au} - \Delta Y_{\pi})^2 \cdot (2Z_{\rm Au} + \Delta Y_{\pi})\label{eq:t}
\end{equation}

\begin{equation}
    ^3{\rm He} = \Tilde{B}_3 \cdot (2N_{\rm Au} - \Delta Y_{\pi}) \cdot (2Z_{\rm Au} + \Delta Y_{\pi})^2\label{eq:He}
\end{equation}

Fig. \ref{fig:eq_d_vs_deltapi} shows the expected dependence of the deuteron, triton and ${}^3$He yield in $4\pi$ (excluding spectators). The red full line shows the expectations for the deuteron, the blue dashed line depicts the tritons and the green dotted line the ${}^3$He. As discussed above, the variation of the isospin leads to a pronounced local maximum of the deuteron production at $\Delta Y_{\pi}=N_\mathrm{Au}-Z_\mathrm{Au}=39$. Also for the tritons and ${}^3$He one observes local maxima, which are symmetric for both cases under the exchange of protons and neutrons. For the triton the maximum is at $\Delta Y_{\pi}=\frac{1}{3}(2N_\mathrm{Au}-4Z_\mathrm{Au})=-\frac{80}{3}$ and for the $^3$He it is located at $\Delta Y_{\pi}=\frac{1}{3}(4N_\mathrm{Au}-2Z_\mathrm{Au})=\frac{314}{3}$. Due to symmetry tritons and $^3$He are equally abundant when isospin is equally distributed between the neutrons and protons at $\Delta Y_{\pi}=39$.

\section{Quantitative estimates}
For the quantitative test of this new idea, we employ the  Ultra-relativistic Quantum Molecular Dynamics (UrQMD) model \cite{Bass:1998ca,Bleicher:1999xi,Bleicher:2022kcu} in version v3.5. UrQMD is based on hadronic and string degrees of freedom and propagates the n-body dynamics via a relativistic QMD implementation. The model includes as explicit degrees of freedom baryonic and mesonic ground states and their resonances up to masses of 4 GeV. Light clusters are produced by phase space coalescence from nucleons at kinetic freeze-out (see \cite{Sombun:2018yqh,Hillmann:2021zgj,Kireyeu:2022qmv} for details).

The collision energy best suited for this study is where the pion/participant ratio is not too much above unity. At such energies, the coupling of the pions and the nucleons is strong and the fluctuations of the net charged pion number has a strong influence on the isospin of the nucleon system. Also the deuteron yield is still comparatively large. Practically, this suggests that Au+Au or Pb+Pb reactions at $\sqrt{s_\mathrm{NN}}=3-8$ GeV are good candidates for this exploration. At lower energies, the probability for large net charged pion fluctuations decreases due to the limited amount of energy, however, the deuteron yield is larger, while at higher energies, the fluctuations are large, however less coupled to the nucleon system and the deuteron multiplicity is lower.

\subsection{Freeze-out time distributions}
To illustrate the idea, we show the freeze-out time distribution of participating nucleons (full blue line), pions (dashed black line), deuterons (dotted red line), and tritons and $^3$He as dotted green and dotted blue lines in central Au+Au collisions at $\sqrt{s_\mathrm{NN}}=3$ GeV in Fig. \ref{fig:freeze-out_time}. One clearly observes that the clusters are formed  after the pions have decoupled from the system. Thus, our expectation that cluster formation proceeds after the emission of the pions is fulfilled. Therefore, cluster formation is sensitive to the isospin fluctuations induced by the preceding pion emission.

\begin{figure}
    \centering
    \includegraphics[width=\columnwidth]{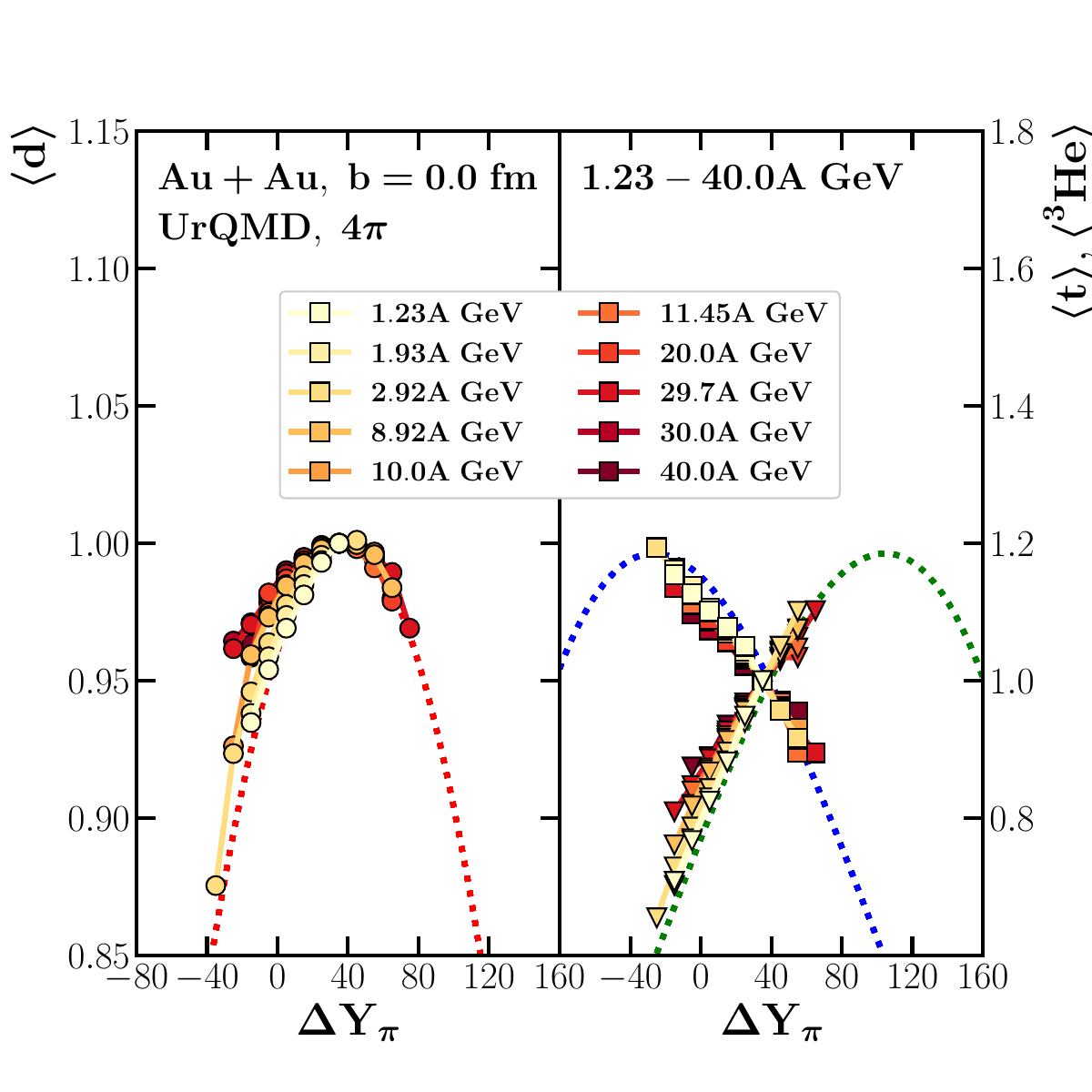}
    \caption{[Color online] Distribution of cluster yields (normalized to unity at $\Delta Y_{\pi}=39$ for better comparison) for various collision energies from $E_\mathrm{lab}=1.23 A\,$GeV to $E_\mathrm{lab}=40 A\,$GeV in ultra-central Au+Au reactions from UrQMD. Left: Deuteron distribution. Right: Triton and ${}^3$He distribution.}
    \label{fig:urqmd_t_he_vs_deltapi_edep}
\end{figure}

\subsection{Light cluster yields versus isospin fluctuation}
We support our estimates with UrQMD simulations in Fig. \ref{fig:urqmd_d_vs_deltapi}. In the left panel we show the deuteron yield as a function of $\Delta Y_{\pi}$ for ultra-central Au+Au reactions at $\sqrt{s_\mathrm{NN}}=3$ GeV. In the right panel, we show the results for ultra-central Au+Au collisions at $\sqrt{s_\mathrm{NN}}=7.7$ GeV. One clearly observes that the simulations at both energies follow our expectations and show a maximum at $\Delta Y_{\pi}=39$.

We also show the results for the heavier triton (blue squares) and ${}^3$He (green triangles) clusters at $\sqrt{s_\mathrm{NN}}=3$ GeV (left panel) and $\sqrt{s_\mathrm{NN}}=7.7$ GeV (right panel) in Fig. \ref{fig:urqmd_t_he_vs_deltapi}. In this case we observe that the maximum for tritons is at $\Delta Y_{\pi}=-26.67$ and $\Delta Y_{\pi}=104.67$ for ${}^3$He. This provides a nice cross check for our proposal as we found the same values from our first estimates.

Finally, Fig. \ref{fig:urqmd_t_he_vs_deltapi_edep} summarizes the results for the energy dependence of the cluster yields as a function of $\Delta Y_{\pi}$ for collision energies from $E_\mathrm{lab}=1.23 A\,$GeV to $E_\mathrm{lab}=40 A\,$GeV in ultra-central Au+Au reactions from UrQMD. In the left part we show the deuteron distribution and in the right part we show the triton and ${}^3$He distributions.  The distribution of cluster yields are normalized to unity at $\Delta Y_{\pi}=39$ for better comparison. We observe that the predicted correlation is present over a wide range of energies.

A detailed inspection of Figs. \ref{fig:urqmd_d_vs_deltapi}-\ref{fig:urqmd_t_he_vs_deltapi_edep} shows that the full UrQMD simulation does not exactly follow the estimates from the simple model presented above. The main reason for this deviation is that in the simple model, isospin is assumed to be shuffled only between nucleons and pions. In the full simulation we see that this assumption is good, but not perfect, because isospin is also shuffled into other hadron species, e.g. $K^+, K^-$, charged $\Sigma$ baryons, ... . This leads to the small discrepancies between the simple model and the full simulation. Let us shortly remark, that the reason, why we suggest to use only the charged pions is because they can be measured rather well on an event-by-event basis.

We conclude that the coalescence model shows a pronounced maximum in the deuteron yield at $\Delta Y_{\pi}=39$ (for Au+Au reactions) and analogous maxima for triton and ${}^3$He. This clearly allows to distinguish coalescence from thermal model calculations, because the thermal model does not  show a dependence\footnote{Let us clarify, that our statements on the thermal model refer to a global grand canonical ensemble as it is most widely used in the field of heavy ion collisions \cite{Cleymans:1992zc,Becattini:1997ii,Florkowski:2001fp,Cleymans:2005xv,Andronic:2010qu,Petran:2013lja,Vovchenko:2015idt,Andronic:2018qqt,Vovchenko:2019pjl}. Of course, in a (micro) canonical ensemble additional correlations might emerge, which may lead to similar effects as observed here for a fully microscopic, energy, momentum and quantum number conserving transport simulation.}
of the cluster yield on $\Delta Y_{\pi}$.

\section{Conclusion}
We proposed to use isospin fluctuations to provide an answer to the question, if light cluster production in heavy ion reactions is by thermal emission at chemical freeze-out or by coalescence at kinetic freeze-out. To this aim, we argued that in a coalescence picture, the light clusters are formed after the pions are emitted, thus the remaining nucleon system is sensitive to the isospin carried away by the pions. This leads to a maximum of the deuteron (and triton and ${}^3$He) yield at a specific $\Delta Y_{\pi}$. In contrast, in a thermal model, the light cluster production is not sensitive to thermal fluctuations of the net charged pion number and the deuteron yield is independent of $\Delta Y_{\pi} = Y(\pi^-) - Y(\pi^+)$.

Only if one would confront coalescence to a fully microcanonical sampler after kinetic freeze out, i.e. after the decays, this comparison could be made. However, as we said before, such a significant extension of the 'standard thermal model' is not what we or most of the community are referring to when speaking about thermal production.

This new method can therefore answer this question on the production mechanism of light clusters. It can be directly probed in Au+Au reactions at $\sqrt{s_\mathrm{NN}}=3$ GeV and $\sqrt{s_\mathrm{NN}}=7.7$ GeV already available at the RHIC-BES or at the CERN-SPS.

\begin{acknowledgements}
This article is part of a project that has received funding from the European Union’s Horizon 2020 research and innovation program under grant agreement STRONG – 2020 - No 824093. Computational resources were provided by the Center for Scientific Computing (CSC) of the Goethe University and the ``Green Cube" at GSI, Darmstadt. This project was supported by the DAAD (PPP Thailand). This research has received funding support from the NSRF via the Program Management Unit for Human Resources \& Institutional Development, Research and Innovation [grant number B16F640076].
\end{acknowledgements}


\end{document}